# Mindscape

## Research of high-information density street environments based on electroencephalogram recording and virtual reality head-mounted simulation

Yijiang Liu[1], Xiangyu Guan[2], Lun Liu[3], Hui Wang[4]
[1,2,4]Tsinghua University [3]Institute of Public Governance, Peking University
[1]liuyijia19@mails.tsinghua.edu.cn [2]guanxy19@mails.tsinghua.edu.cn
[3]liu.lun@pku.edu.cn [4]wh-sa@mail.tsinghua.edu.cn

*This study aims to investigate, through neuroscientific methods, the effects of particular architectural elements on pedestrian spatial cognition and experience in the analysis and design of walking street spaces. More precisely, this paper will describe the impact of the density variation of storefront signs on the brainwaves of passersby in East Asian city walking streets, providing strategies and guidelines for urban development and renewal. Firstly, the paper summarizes the research method through the review of research questions and related literature; secondly, the paper establishes experiments via this path, analyzing results and indicators through data processing; finally, suggestions for future pedestrian street design are proposed based on research and analysis results.*

**Keywords:** Urban Studies, Neuroscience, EEG, Street Information Density
**Acknowledgement:** *This work is supported by National Natural Science Foundation of China (52378022); This paper has been accepted at eCAADe 2024 Conference.*

## INTRODUCTION

In the development of East Asian cities, signs as an architectural element have always been an integral medium for information dissemination in neighborhoods, and due to the slow-moving characteristics and high commercial value given by the gathering of young consumer groups, walking districts have become significant distribution places for signs. In pedestrian areas, signs and the graphical and textual information they convey merge with the street space, together forming the unique street temperament of East Asian cities with a high density of information.

Today, with the advancement of medical technology, people can use brain-computer interfaces (BCI) to directly obtain real-time brain data on spatial cognition and experience and monitor and analyze cognitive conditions accordingly. Moreover, the popularity of interactive technologies such as virtual reality (VR) allows scientists to create interactive virtual environments (VE), which are closer to reality, and to grasp the precision of research problems by accurate parameter adjustment, controlling stable research variables, and excluding irrelevant variables. This paper discusses research based on new interactive technologies and new human perception data to help urban designers make decisions and guidelines. Especially under the influence of different densities of street signs, how people's feelings change with the changes in sign density is investigated. We built digital models based on the common composition forms of pedestrian streets in East Asia and used parametric tools to determine the placement layout of storefront signs under different information density gradients to simulate real street scenes. We obtained controllable storefront sign textures through training generative artificial intelligence, resulting in the virtual environment for the experiment. Participants were asked to act as customers walking through commercial streets in a virtual environment, touring specified paths, and monitored in real-time for brainwaves and visual fields through worn electroencephalogram (EEG) and VR devices to track changes in brain activity; questionnaires and post-experimental interviews were used to gain a deeper understanding of the overall mental state of the participants. After analyzing and validating experimental data, associations between normalized EEG information indicators and information density were obtained.

## RESEARCH THEME & REVIEW

The study encompasses the investigation of three themes: the creation of immersive virtual environments and the application of neuroscientific methods to the study of spatial cognition. It examines the simulation of real-world scenarios within immersive virtual environments and the control of variables. Additionally, it explores the quantitative research and analysis of cognitive spatial perception in pedestrian street spaces using neuroscientific methods. Researchers utilize a variety of technologies to create walking environments that closely resemble reality, and through data analysis methods combined with classical psychological approaches, they collectively verify users' perceptions.

## METHODOLOGY

### Variable

Existing research has preliminarily addressed the feasibility of using VR as a substitute for real-life scenarios in academic studies. For the purpose of controlling variables, we have established methods in VR simulation and EEG monitoring. However, whether the specific research variables chosen for urban problems in this study are rational, and whether the corresponding virtual environments (VEs) meet the needs of immersion, have not yet been scientifically validated. To address these issues, the experimental group set up a pre-experiment focusing on the selection of different elements.

In the pre-experiment, we prepared a VR video recorded on a typical real commercial street and a 3D model of the street modified by the experimental group (Figure 1). Without informing the research theme, subjects were asked to wear VR headsets and experience a two-minute virtual street walking task. Thirteen subjects were recruited for the pre-experiment, and each was required to self-report and complete questionnaires after the experience. The questionnaire consists of two parts: the first part is semi-open-ended questions about the real street scene to confirm the rationality of variable selection, including two questions: "1. In the VR video's virtual street experience, what do you think is the most significant street element? 2. Recall the walking experience on the real street. In the VR video's virtual street experience, what do you think is the biggest difference from the real street?" Each question includes options such as street noise, paving, storefront signs, vegetation, pedestrians, ground litter, etc., and users are required to select 1-3 items that match the problem description while being encouraged to provide free-form answers. The second part is a structured questionnaire based on the presence questionnaire published by Slater in 2009 to verify the feasibility of the virtual scenes assembled by the experimental group for use in experiments.

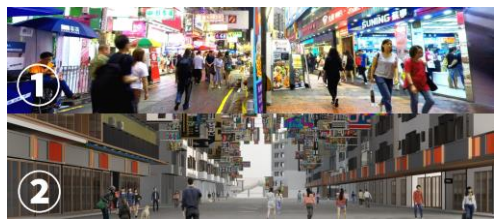

Figure 1
Pre-experimental Scene Schematic Diagram

The results of the first part of the questionnaire showed the rationality of variable selection: the frequency of the architectural storefront sign element in the results of question one was the highest (10/13), significantly higher than the second place (5/13). Three subjects supplemented key words such as lighting and neon lights in free-form answers. This result proves that our focus on storefront signs as the main independent variable of street information density is fully rational; The results of question two showed that more than half of the subjects were particularly sensitive to street noise elements, noting their absence in the sole VR headset experience.

The structured questionnaire part showed the success of creating a sense of immersion: the final score was significantly higher than the reference score required for immersion in the presence questionnaire. Concurrently with the investigation of differences from real streets, the final formal experiment decided to select suitable commercial street white noise to accompany the virtual street walking in the VR.

## EEG

**Spatial Features.** In the application fields of neuroscience and cognitive science, the brain can be roughly divided into five regions—frontal lobe, parietal lobe, temporal lobe (including left and right temporal lobes), and occipital lobe—to better understand the structure and function of the brain. The amplitude of EEG signals represents the intensity of the potential, thereby reflecting the activity of the corresponding brain regions. This study used EEG monitoring tools and corresponding mathematical relevance analysis methods to discern, as the scene switches continuously and the street information density increases, the most active brain regions and points of subjects, as well as the brain regions and points most sensitive to such changes.

**Frequency Domain Features.** The study calculated three indicators based on EEG frequency band analysis methods: arousal, cognitive load, and focus degree, thereby analyzing the emotional cognitive state of people in experimental scenes. The experiment first transformed the original EEG into discrete frequency bands via Fast Fourier Transform (FFT), and all indicators were built on a non-linear weighted compound model of frequency bands and electrodes. To better compare emotional indicators, each participant was required to perform a benchmarking exercise to normalize their indicator data.

Arousal is a response to the individual's instant emotional intensity and is highly correlated with the $\alpha$ and $\beta$ bands in the frontal lobe area; Cognitive load represents the brain's processing burden and memory needs. Scores range from 0 to 100. Generally, the "optimum" score range is between 55 and 75, indicating that the brain is active and information is understood and remembered. Lastly, *beta/(alpha + theta)* was used to calculate subject Focus, mainly targeting brain regions responsible for managing learning, mental state, and attention.

## SAM Value

SAM scale, or Self-Assessment Manikin (SAM) questionnaire, is a widely recognized emotional state survey scale that assesses the user's mental state by measuring three independent indicators (pleasure-displeasure, degree of arousal, dominance-submissiveness). For subjects with healthy cognitive levels and language abilities, it is a good supplement to EEG data. The experiment will use SAM scores to aid the evaluation of EEG data for a comprehensive judgment of the subject's emotional state.

## VIRTUAL ENVIRONMENT

To pursue the authenticity of the virtual environment and obtain more accurate cognitive data, the main body of the experimental scene chose commercial street architectural modules with rich architectural details purchased from online model libraries. The architectural modules were combined to ensure properties, such as the facade of buildings and the ratio of street width to height, in the virtual environment were homogeneous and stable.

After perfecting the basic scene, researchers defined the main research object: the total area of storefront signs passed per unit advancement distance in the virtual environment as the sign density, and approximated sign density to information for measurement, setting the calculation rules for density (Figure 2). Storefront signs were generated on the surfaces of the building modules on both sides of the walking street through the collaboration of modeling software Rhino and parametric design plugin Grasshopper, with specific points generated, the specific process as follows: Firstly, the walking street was sliced along the direction of travel into equally distanced layers, with the intersection lines of each layer and building modules serving as the sign point generation positions; then, according to the number of points input, a random number of horizontal signs with random sizes and aspect ratios between 1:3 to 1:4, or one larger vertical sign, were generated at the intersection line.

This generation program allows users to adjust sign density by altering layer spacing, the number of points per layer, etc. The process referred to street sign regulations in cities such as Hong Kong and extracted real parameters to ensure scenario and real-world problem matching, such as the distance between layers should be greater than 2.4 meters, and a single sign's width should not exceed 4.2 meters. Under such restrictions, there exists a maximum value for sign density, which the experimental group defined as 100%, and correspondingly, a density of 0% when there were no storefront signs at all, allowing the experimental group to build a virtual street environment with changing sign density.

To cover a higher precision range of densities, enabling multiple repeated tests to reduce errors; and prevent users from experiencing emotional breakpoints due to drastic differences before and after in continuous scenes, the researchers divided the virtual scenes into five groups based on the change of sign density from 0% to 100%. Each group contained density changes within a narrower range within their area, displaying macro changes in sign density with larger gradients between groups.

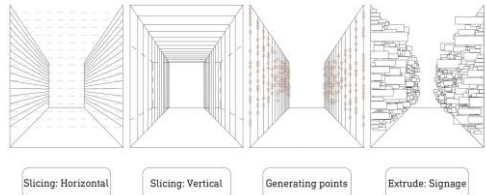

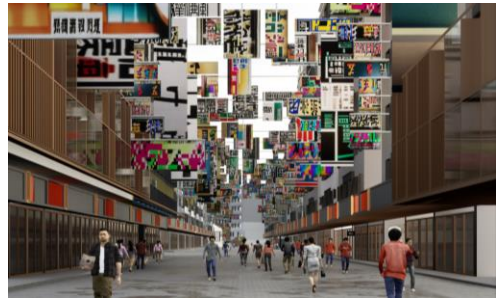

Figure 2 Partition the studied space while defining density, and generate signs accordingly.

Figure 3 Experimental Street Scene in Virtual Reality Headset, Displaying a Specific Density

At the same time, considering that storefront signs primarily contain textual information, and the content of the graphics combined with vibrant color schemes significantly affect the visual experience during street walking, pedestrians are often attracted by the position, colors of the signs and spend energy understanding textual information. In this study, to focus on the research question on density and to potentially avoid the inconsistency in cognitive loss during the process of understanding the content of the signs, researchers obtained controllable sign textures through generative artificial intelligence. Visually similar to real signs, but without corresponding content information (Figure 3).

To achieve the above, researchers collected a large number of billboards with Chinese as the base from the internet, classified and processed them based on the visual complexity of the graphic surfaces, and used them as the image dataset. Three Lora models were trained on the Stable Diffusion platform, corresponding to three different levels of graphic complexity. Also, for the purpose of controllable sign content, the study only selected one Lora model closest to real storefront images in the end and generated 215 visually similar pseudo-sign images based on it (Figure 4).

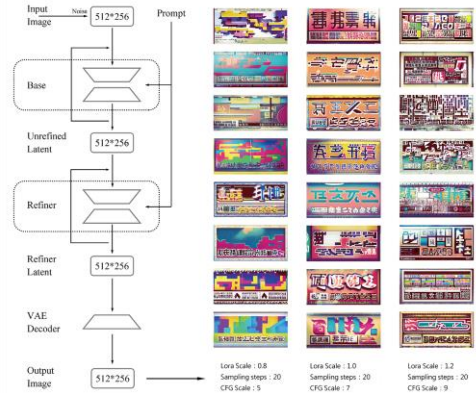

Figure 4
Texture Generate. The Process for Training and Generating storefront texture in Stable Diffusion

Following the exploration of information density in the above experiments, researchers attempted to further research typical high-density scenarios: commercial streets, as typical research subjects, have different building attributes such as street ratio, vegetation greening rate, and the horizontal and vertical arrangement of signs. In the subsequent experiments with a higher baseline density, researchers included other street properties such as street ratio, vegetation greening rate, and sign color tone in the experiment variables, asking subjects to consider these changes after experiencing density change experiments as countermeasures against high information density. In this experiment, scenes adopted a higher degree of categorical blurring: all street properties studied were divided into three gradient changes. The above classification tends to let users subjectively choose the scenarios with better counteracting effects, which shows the group trends and preferences of users within a larger range and can provide preliminary guidelines for future street design.

## EXPERIMENT SETUP

The virtual scenes were processed through Unreal Engine 5 for VR scene cognition experiments, with five groups of virtual scenes loaded into Unreal Engine for interaction and VR experience processing: we modified the character movement from teleportation to normal movement and set a specific virtual environment experience route. This action hopes to control the uniformity of the subject's field of vision as much as possible while ensuring immersion in the experiment to reduce errors.

The VR device used in the experiment is the Oculus Quest 2 produced by Meta, which ensures the smoothness and stability of the image through usb3.0 wired direct connection. The EEG research uses the 14-electrode non-invasive EEG device EPOC X developed by Emotiv, and the software emotiv pro accompanying the device for preliminary data processing and analysis. The device uses wet electrode sampling, 14 sampling electrodes based on the international 10-20 positioning system, distributed in the frontal lobe (AF3, AF4, F3, F4, F7, F8), central frontal lobe (FC5, FC6), occipital lobe (O1, O2), parietal lobe (P7, P8), and temporal parts (T7, T8). The device sampling frequency is 128SPS. Emotiv pro software supports EEG signal recording, export, and preliminary analysis during experiments. The data exported by this EEG device has been verified against medical-grade head-mounted devices; apart from relatively high baseline noise, it can provide accurate EEG data collection.

The experiment required subjects to wear both VR and EEG devices at the same time. To avoid signal quality degradation due to body movement, particularly significant head shaking, and to reduce experimental baseline noise, subjects were asked to sit and maintain a fixed posture, replacing body movement as much as possible with changes in their line of sight during scene experience. 22 students from Tsinghua University were recruited as subjects for the experiment, including 6 females and 16 males, with an average age of 21.8 years. All subjects were provided with certain financial subsidies for participation (Figure 5).

## EXPERIMENT PROCEDURE

All subjects in the experiment were arranged to experience from low to high density to eliminate potential errors caused by different viewing orders of different subjects. Subjects were first required to wear VR glasses and EEG devices and adjust to ensure the field of vision was clear and stable, and the correct EEG signal point position and good contact were established. All subjects were asked to close their eyes and rest for 30 seconds, imagining themselves shopping on a commercial street, then open their eyes to start running the VR street environment and record EEG information. The experiment required subjects to actively pay attention to street information and maintain visual focus. Each scene required subjects to move forward for about 60 seconds in the same direction at an even pace, after which the subjects removed the VR device and stopped the EEG data recording. After each scene, subjects needed to complete the SAM scale based on their overall experience of the current scene. After completion, the subjects would continue to close their eyes and rest for 60 seconds before entering the next scene.

After completing the EEG signal collection and subjective scale report under the density change scene, subjects had a rest for three minutes and then assisted us in completing the iterative design of the street scene. Users provided optimization judgment results after viewing experimental scenes with changes in street properties such as street ratio, greening rate, and sign color tone as an important reference for subsequent design guidelines.

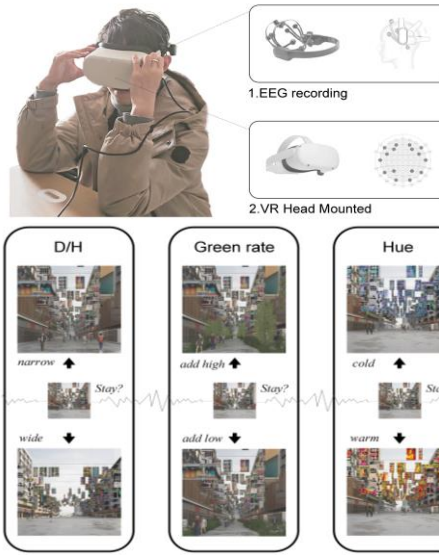

Figure 5 Experimental Device

Figure 6 Street attributes options presented to participants for as countermeasures against high information density streets

Subjects were placed in high-density environments and were asked to adjust the above three variables in the VR scene based on their comfort level. Researchers recorded each person's selection tendency and used the results for street scene preference analysis (Figure 6).

## ANALYSIS

The preprocessing process is carried out using the EEGlab toolbox on the MATLAB R2022b platform, where a 0.1~70 Hz band-pass filter is deployed to isolate relevant frequency bands. Data from each epoch is extracted within a 2-second time window (with 0.5s overlap) and utilizes Independent Component Analysis (ICA) to remove artifacts such as blinking and electromyographic signals.

Figure 7
EEG electrode site correlation analysis. Through a comprehensive correlation analysis of 14 electrode sites

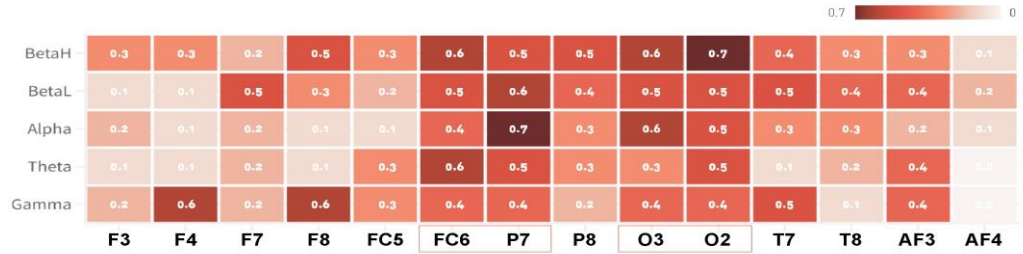

**Brain region and electrode sites.** As shown, the five EEG points with the highest correlation to the variability of street information density are P7, O1, O2, T7, and FC6. In terms of brain regions, the strongest responses are in the parietal and occipital lobes (r=0.5-0.6), indicating that besides processing visual information, these areas also have spatial perception and localization functions. Additionally, the temporal (T7) and frontal (FC6) lobes also show a certain statistical correlation (r>0.4), suggesting that apart from basic language and motor planning functions, they also influence spatial cognition. In terms of frequency domain data, the three bands with the highest correlations are the α, β, and θ waves (Figure 7).

**Emotion indicators.** The study plots the emotional responses of subjects under different information densities both locally and globally, using a moving average with a period of 30 to reveal long-term trends (Figure 8). Average results from 22 subjects show that users' focus tends to decrease overall as street information density increases. In scenes one and two, a slight upward trend in focus is evident, implying that a suitable amount of information density can promote mental focus during walking.

However, as the experiment progresses to the transition between scenes three and four (information density=55%), the data show that focus has reached its lowest point. As density continues to increase, focus levels maintain around 0.4.

Participants' arousal levels initially increase then decrease, showing a slight upward trend overall. Arousal is at a very low base level when density is 0, gradually increasing and reaching a local peak at scene two (density=25%). The global peak occurs in scene three (density=50%). During the subsequent decrease in arousal, a local peak appears again at a density of 80%, which the researchers speculate correlates to the arousal value of negative emotions. Furthermore, when density reaches its maximum value, arousal returns to a minimum value close to that at density 0; this is speculated to be due to the excessive and complex street information density, making it difficult to extract practical information, causing a numbing of user psychological experience.

The cognitive load on users slightly increases overall. Data show that at the beginning of the experimentation with increasing density, cognitive load swiftly rises, reaching 80 points at a density of 25%, nearly the overall peak, indicating an average information overload for participants at this point. As subjects gradually adapt to high-density scenes, this emotional indicator drops and stabilizes around 50, with peaks appearing at densities of 65% and 80%, and ultimately showing a downward trend.

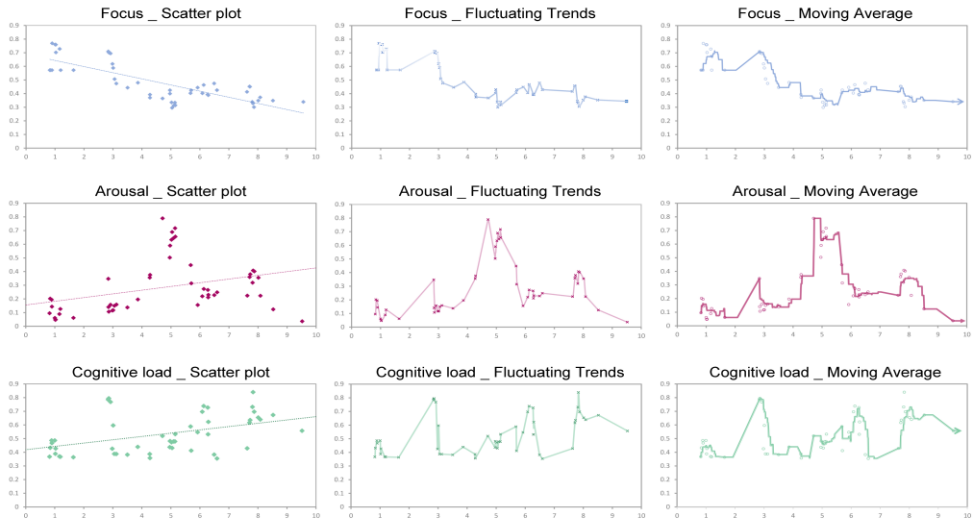

Figure 8 Visualization of Participants' Emotional Responses and trends to changes in street information density

Researchers analyzed the causes of the two drops, considering one to be due to the continued sharp increase in density from an already high level, causing a loss in processing ability; the other due to a certain homogeneity in the five experimental scenes in terms of street layout and shop sign content, leading to an adaptation by the subjects.

**SAM.** The SAM questionnaire assessed the subjective cognitive situations of all 22 subjects, showing trends such as a decrease in the pleasure-unpleasure index as density gradually increased, with cyclical peaks between 20%-40% and 60%-80%; the degree of Arousal showed an increasing then decreasing trend, with a significant peak between 60%-80%; dominance-submissiveness logically decreased with increasing density. These indicators, after correlation study with processed EEG data, revealed some matching trends (Figure 9).

The chart presents a correlational study between SAM questionnaire data and EEG measurement analysis values, displaying a strong negative correlation between the Arousal and Attention indicators in the SAM questionnaire, a strong positive correlation with the Cognitive Load indicator, and a strong positive relationship between Dominance and Attention. However, it is noteworthy that, while both measurement methods show similar trends in Arousal, there is a significant lag evident. This lag could be the reason behind the less obvious correlation observed.

**Variable selection.** In the second phase of the experiment, subjects selected changes among several scene attributes. In choosing street aspect ratio attributes, nearly 80% of subjects chose to change the current scene's aspect ratio to address excessively high information density, yet the overall direction of change was unclear. 45.4% of subjects preferred a larger aspect ratio to alleviate the tension brought by high information density with more spacious streets; meanwhile, 31.8% chose a smaller aspect ratio, reporting that narrow streets created a more encompassing and compact atmosphere, making their unique characteristics more appreciated. Similarly, nearly 90% of subjects opted for changes in vegetation coverage rate attributes, with 63.6% voting for scenarios with lower greenery rates and 27.3% for higher rates. Compared to extensive coverage, moderate greenery is more favored in high-density scenes; more than half of the subjects (54.5%) opted to adjust the existing shop sign images towards warmer color tones, while fewer (18.2%) chose cooler tones over maintaining the status quo (Figure 10).

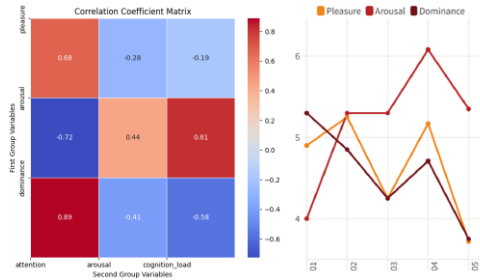

Figure 9 Participants' Self-Assessment Manikin (SAM) Scale Results and Correlation with EEG Indices

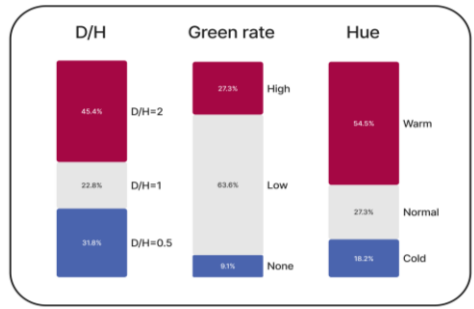

Figure 10 Participants' selection results for street attributes as countermeasures

## DISCUSSION & RESULTS

The results indicate that the EEG points most correlated with changes in street information density are P7, O1, and O2, located in the parietal and occipital lobes responsible for visual processing and spatial cognition. This confirms the experiment's independent variables can effectively stimulate relevant brain areas, demonstrating the feasibility and extensibility of this research method.

It's also shown that as information density increases, participants' focus decreases gradually, while arousal and cognitive load exhibit cyclical peaks. Notably, the three peaks in arousal and cognitive load coincide at densities of 25%, about 55%, and 80%, though cognitive load shows a slight lag compared to arousal at medium density levels (46%-50%), presenting a high arousal and low load emotional state. Such scenes are advocated for in design.

The study additionally reveals that when street information density reaches excessively high levels, users prefer suitable, non-obstructive greenery to ease discomfort, while the choice of street width does not show significant differences. Interestingly, against the backdrop of frequently suggested warm color storefront signs like red and yellow in urban planning worldwide, nearly half of the subjects still chose cool or the original random colors (though warm colors were in the majority). This suggests that urban street planners should consider a reasonable mix of diverse color palettes on the basis of predominantly warm tones.

Furthermore, to focus on the information density variable, the study didn't change attributes like main street body, background scenery, and walking paths, which also led to subjects prematurely recognizing some subsequent scenes to some extent. Based on the proven effectiveness of this research method, future studies could explore more varied architectural scenes.

In summary, we can conclude that changes in street information density elicit the strongest responses in the parietal (P7) and occipital (O1, O2) lobes and that moderate information density in street scenes can stimulate high arousal while maintaining a healthy cognitive load in pedestrians. Expanding to real commercial street planning, designers can refer to the permitted threshold of street signs density and consider factors like greenery and color coordination to plan suitable street spaces, offering pedestrians information-rich environments to maintain positive exploration and psychological states.